\documentclass[12pt]{article}
\usepackage{epsf}
\usepackage{amsmath}
\usepackage{graphics}
\usepackage{cite}

\renewcommand{\thefootnote}{\#\arabic{footnote}}
\begin{document}

\newcommand{\gtrsim}{ \mathop{}_{\textstyle \sim}^{\textstyle >} }
\newcommand{\lesssim}{ \mathop{}_{\textstyle \sim}^{\textstyle <} }

\newcommand{\rem}[1]{{\bf #1}}

\renewcommand{\thefootnote}{\fnsymbol{footnote}}
\setcounter{footnote}{0}
\begin{titlepage}

\def\thefootnote{\fnsymbol{footnote}}

\begin{center}
\hfill November 2016\\
\vskip .5in
\bigskip
\bigskip
{\Large \bf Cyclic Period in the CBE Model}

\vskip .45in

{\bf Paul H. Frampton\footnote{email:paul.h.frampton@gmail.com;homepage:paulframpton.org}}

{\em University of Salento, Lecce, Italy}

\end{center}

\vskip .4in
\begin{abstract}
In a cyclic entropy model in which the extroverse is jettisoned at
turnaround with a Come Back Empty (CBE) assumption,
we address matching of the contaction scale factor 
$\hat{a}(t)=f(t_T){a}(t)$ to the expansion scale factor $a(t)$, where $f(t_T)$ 
is the ratio at turnaround of the introverse to
extroverse radii. Such matching is necessary for infinite cyclicity
and fixes the CBE period at $\sim 2.6Ty$. 
\end{abstract}
\end{titlepage}

\renewcommand{\thepage}{\arabic{page}}
\setcounter{page}{1}
\renewcommand{\thefootnote}{\#\arabic{footnote}}

\newpage

\section{Introduction to the CBE model}

In physics there exist simple-to-state questions which are 
difficult to answer. For example, here is a 
plausible examination question: Show how to construct an infinitely cyclic cosmological model which is consistent with the second law of thermodynamics.

\bigskip

\noindent
This has been studied since 1931 when a no-go theorem
\cite{Tolman} of Tolman stated that, subject to certain assumptions,
there cannot exist any solution to creating
the cosmological sequence 
\begin{equation}
Expansion \longrightarrow Turnaround \longrightarrow
Contraction \longrightarrow Bounce \longrightarrow~~~ etc.
\label{CBEmodel}
\end{equation} 
where the entropy of the universe obeys the second law of thermodynamics.
Fortunately in this case,
one can be guided by a more recent discovery about Nature.

\bigskip

\noindent
The most important questionable assumption implicit in the no-go theorem
\cite{Tolman} of 1931 was pointed out not by a physicist but 
by Nature herself in 1998 when observers
discovered \cite{Perlmutter,Riess} that the expansion rate of the universe
is accelerating. To my knowledge,
nobody had questioned prior to 1998 that the expansion rate
was decelerating. Once one knows that it is accelerating it is very enlightening
with respect to the second law of thermodynamics because the superluminal
accelerated expansion creates, starting at the onset 
of dark energy domination, the extroverse into which entropy built up from
irreversible processes during expansion can be jettisoned with impunity from
a retained introverse. The CBE (Comes Back Empty)
assumption is that the retained introverse
contains energy of radiation, dark energy and curvature but no matter, 
luminous or dark, including no black holes.

\bigskip

\noindent
The superluminal accelerated expansion is surely the most important
discovery in observational cosmology since Hubble\cite{Hubble} 
established the expansion of the universe. In terms of entropy it
made it possible to distinguish the two parts of the universe after
the dark energy domination began at $t_{DE} \sim 9.8$Gy.
Those two parts which play a crucial role in the CBE model
are the introverse and extroverse which we shall now define.
Note that although the CBE assumption was first introduced in
\cite{BF,FramptonBook} the only CBE model discussed in the
present paper is the recently improved CBE model \cite{Frampton} 
which eschews the use of phantom dark energy.

\bigskip

\noindent
The introverse is the same as the visible universe, or particle horizon,
whose radius $R_{IV}(t)$
is given by
\begin{equation}
R_{IV}(t) = c \int_0^t \frac{dt}{a(t)},
\label{IV}
\end{equation}
where $a(t)$ characterizes the expansion history of the universe, being the
scale factor in the FRLW metric which assumes homogeneity and isotropy
\begin{equation}
ds^2 = dt^2 - a(t)^2 \left[ \frac{dr^2}{1-k(t)r^2} + r^2 (d\theta^2 + \sin^2\theta d\phi^2) \right],
\end{equation}
where $k(t)$ is the curvature.

\bigskip

\noindent
Inserting the well-known expansion history and normalizing the scale factor to
$a(t_0) = 1$ at the present time $t_0 = 13.8Gy$ (all times are measured relative
to the would-have-been big bang) one finds that at the end of the radiation-dominated
era $a(t_m=47ky)=2.1\times 10^{-4}$, a value which will be important in the CBE
matching
condition to be discussed in this paper.  At the commencement of the superluminal accelerated expansion
at $t_{DE}=9.8Gy$ one finds $a(t_{DE})=0.75$. After this time the scale factor 
($t\geq9.8Gy$) is
\begin{equation}
a(t) = 0.75 \exp [H_0 (t - t_{DE})],
\label{acceleration}
\end{equation}
where the observed value of the Hubble constant is $H_0\simeq(13.8Gy)^{-1}$.

\bigskip

\noindent
Substituting Eq.(\ref{acceleration}) in Eq.(\ref{IV}) one finds that the radius
of the introverse $R_{IV}(t)$ grows from an initial value at $t=t_{DE}$
\begin{equation}
R_{IV}(t_{DE}) = 39Gly,
\end{equation}
to its present value
\begin{equation}
R_{IV}(t_0)=44Gly,
\label{presentIV}
\end{equation}
and reaches an asymptotic value at $t\sim 50Gy$ so that
\begin{equation}
R_{IV}(50Gy \leq t < \infty) \simeq 58Gly.
\label{asymptoticIV}
\end{equation}

\bigskip

\noindent
The extroverse starts to form after $t=t_{DE}$ and its radius $R_{EV}(t)$ is
defined initially as equal to the intoverse radius
\begin{equation}
R_{EV}(t_{DE}) = R_{IV}(t_{DE}) = 39 Gly,
\end{equation}
and thereafter for $t_{DE} \leq t$ is
\begin{equation}
R_{EV}(t) = \frac{a(t)}{a(t_{DE})}R_{EV}(t_{DE}).
\end{equation}

\bigskip

\noindent
This leads to the present value of the extroverse radius
\begin{equation}
R_{EV}(t_0)=52Gly,
\label{presentEV}
\end{equation}
which is significantly above $R_{IV}(t_0)$ given by Eq.(\ref{presentIV}), as discussed in
\cite{Frampton}.

\bigskip

\noindent
Future values of $R_{EV}(t)$ can be illustrated by examples.
When the introverse radius approaches its asymptotic value, the extroverse radius is already much larger than $R_{IV}(t=50Gy)$
given by Eq.(\ref{asymptoticIV}) namely
\begin{equation}
R_{EV}(t=50Gy) = 720Gly.
\end{equation}
An interesting later value of $R_{EV}(t)$, 
extraordinarily large, is
\begin{equation}
R_{EV}(t=1Ty) = 5.6\times10^{32} Gly,
\end{equation}

\bigskip

\noindent
In the CBE model, at a turnaround time $t=t_T$ to be determined in the
next subsection, the scale factor for the contracting universe is
$\hat{a}(t) = f(t_T)a(t)$ with the fraction $f(t_T)$  given by
\begin{equation}
f(t_T) = \frac{R_{IV}(t_T)}{R_{EV}(t_T)}.
\label{fractionf}
\end{equation}

\bigskip

\noindent 
As shown in \cite{Frampton}, this reduction in size of the adiabatically
contracting universe with low entropy explains, without any need
for an inflationary era, the flatness
observed for the present universe and further predicts that the present flatness
is accurate to many decimal places.
 
\bigskip
\bigskip

\section{Matching of the Scale Factor}

\bigskip

\noindent
An important requirement for infinite cyclicity is that the scale factor $a(t)$
for the expansion era be matched correctly to that of the previous contracting
era. Recall the the scale factor is redefined as $\hat{a}(t) = f(t_T) a(t)$ at
turnaround so that, at first sight, it might appear that an inverse transformation
$a(t) = f(t_T)^{-1}\hat{a}(t)$ might be necessary. However, this is not the case
because the subluminal decelerating contraction rate is far more gradual than
the superluminal accelerating expansion rate.

\bigskip

\noindent
The contraction is radiation dominated throughout so that the relevant matching
condition is at the transition time, $t_m \sim 47 ky$, between radiation domination and matter domination
of the expansion era, namely
\begin{equation}
\hat{a}(t_m) = a(t_m) = 2.1\times 10^{-4}.
\label{matching}
\end{equation}

\bigskip

\noindent
This matching condition allow us to fix the turnaround time $t_T$ of the CBE model
and hence its cyclic period, as follows.

\bigskip

\noindent
First note that $t_T$ is necessarily in the asymptotic region of the introverse
where 
\begin{equation}
R_{IV}(t_T) \simeq 58 Gly,
\end{equation}
and consequently
\begin{equation}
\hat{a}(t_T) = f(t_T) a(t_T) = \frac{58Gly}{R_{EV}(t_T)} a(t_T).
\label{ahatturnaround}
\end{equation}

\bigskip

\noindent
We know also that
\begin{equation}
R_{EV}(t_T) = a(t_T) R_{EV}(t_0) = a(t_T) \times 52 Gly,
\end{equation}
which, when combined with Eq.(\ref{ahatturnaround}), reveals that
\begin{equation}
\hat{a}(t_T) = \frac{58 Gly}{52 Gly} = 1.11,
\label{ahat}
\end{equation}
independent of the turnaround time $t_T$ provided that it is in the asymptotic
region $t_T \gtrsim50 Gy$.

\bigskip

\noindent
The matching condition, Eq.(\ref{matching}) is now straightforward to implement
because $\hat{a}(t)$ contracts with the radiation-dominated behavior
\begin{equation}
\hat{a}(t) = \hat{a}(t_T) \left( \frac{t}{t_T} \right)^{\frac{1}{2}},
\end{equation}
and the matching requirement is therefore
\begin{equation}
\hat{a}(t_m) = 1.11 \left( \frac{47 ky}{t_T} \right)^{\frac{1}{2}}  = a(t_m) = 2.1\times 10^{-4},
\end{equation}
which has the unique solution $t_T = 1.3 Ty$.

\bigskip

\noindent
Only with this choice of turnaround time does the contracting universe match smoothly
on to the time-reverse of the expansion radiation dominated era in such a manner that
infinite cyclicity is achieved. The total cyclic period of the CBE model is
thus
\begin{equation}
\tau_{CBE} = 2 t_T = 2.6 Ty.
\end{equation}

\bigskip
\bigskip

\section{Discussion}

\bigskip

\noindent
In the CBE model, gravitational interactions play a role in the overall
behavior of the expansion and contraction of the introverse but not
in its entropy because the gravitational entropy which is dominated by
black holes is jettisoned at turnaround to the extroverse. 

\bigskip

\noindent
The model solves the question of constructing a cyclic
cosmological model which respects the second law of thermodynamics
and it seems unlikely that such a difficult and highly constrained
question can have two really different solutions.

\bigskip
\bigskip

\section*{Acknowledgement}
Thanks are due for useful discussions with F. Englert,
J.M. Maldacena, R. Penrose and J.C. Taylor.

\end{document}